\documentclass[conference]{IEEEtran}
\IEEEoverridecommandlockouts
\usepackage[inline]{enumitem}
\usepackage{graphicx}
\usepackage{xcolor}
\usepackage{csquotes}\usepackage{tikz}
\usetikzlibrary{shapes.geometric}
\usepackage[colorinlistoftodos, bordercolor=yellow, disable]{todonotes}
\makeatletter \@ifpackagewith{todonotes}{disable}{}{
\paperwidth=\dimexpr \paperwidth + 6cm\relax
\oddsidemargin=\dimexpr\oddsidemargin + 3cm\relax
\evensidemargin=\dimexpr\evensidemargin + 3cm\relax
\marginparwidth=\dimexpr \marginparwidth + 3cm\relax
}
\makeatother
\hypersetup{colorlinks=true, linkcolor=black, citecolor=black, filecolor=black, urlcolor=black}
\usepackage[backend=biber, maxcitenames=1, url=false, doi=false, isbn=false]{biblatex}
\newcommand{\ie}{i.e.\ }
\newcommand{\eg}{e.g.,\ }
\begin{document}

\title{Automotive Cybersecurity: Foundations for Next-Generation Vehicles}
\author{
\IEEEauthorblockN{Michele Scalas, \IEEEmembership{Student Member,~IEEE}}
\IEEEauthorblockA{\emph{Department of Electrical and Electronic Engineering}}
\\
\textit{University of Cagliari}\\
Cagliari, Italy\\
michele.scalas@unica.it
\and
\IEEEauthorblockN{Giorgio Giacinto, \IEEEmembership{Senior Member,~IEEE}}
\IEEEauthorblockA{\textit{Department of Electrical and Electronic Engineering}}
\\
\textit{University of Cagliari}\\
Cagliari, Italy\\
giacinto@unica.it
}

\maketitle

\begin{abstract}
The automotive industry is experiencing a serious transformation due to a digitalisation process and the transition to the new paradigm of \emph{Mobility-as-a-Service}. The next-generation vehicles are going to be very complex cyber-physical systems, whose design must be reinvented to fulfil the increasing demand of \emph{smart} services, both for safety and entertainment purposes, causing the manufacturers' model to converge towards that of IT companies. \emph{Connected cars} and \emph{autonomous driving} are the preeminent factors that drive along this route, and they cause the necessity of a new design to address the emerging cybersecurity issues: the \textit{"old"} automotive architecture relied on a single closed network, with no external communications; modern vehicles are going to be always connected indeed, which means the \emph{attack surface} will be much more extended. The result is the need for a paradigm shift towards a \emph{secure-by-design} approach.

In this paper, we propose a systematisation of knowledge about the core cybersecurity aspects to consider when designing a modern car. The major focus is pointed on the in-vehicle network, including its requirements, the current most used protocols and their vulnerabilities. Moreover, starting from the attackers' goals and strategies, we outline the proposed solutions and the main projects towards secure architectures. In this way, we aim to provide the foundations for more targeted analyses about the security impact of autonomous driving and connected cars.
\end{abstract}

 \section{Introduction}
\label{sec:1}

\IEEEPARstart{T}{he} automotive industry is experiencing a serious transformation due to a digitalisation process in many of its aspects and the new mobility models. A recent report by \textcite{PwC2018} states that by 2030 the vehicle parc in Europe and USA will slightly decline, but at the same time the global industry profit will significantly grow.
The main factor for this phenomenon is the concept of \emph{Mobility-as-a-Service} (MaaS), \ie the transition to car sharing and similar services, at the expense of individual car ownership (expected to drop from 90\% to 52\% in China~\cite{PwC2018}). In this sense, the main keywords that will contribute to this new model are \emph{'connected cars'} and \emph{'autonomous driving'}. 
\newline
According to \textcite{UpstreamSecurity2018}, by 2025 the totality of new cars will be shipped connected, intending as connected not only the possibility of leveraging Internet or localisation services but the adoption of the V2X (Vehicle-to-\textit{X}) paradigm. This term refers to the capability of the car to communicate and exchange data with other vehicles (V2V, Vehicle-to-Vehicle), with a generic infrastructure (V2I) or with pedestrians (V2P). The typical application of these models is smart cities, with the aim of optimising traffic management, sending alerts in case of incidents, coordinating a fleet of vehicles.
\newline
As regards autonomous driving, it consists in expanding the current Advanced Driver Assistance Systems (ADASs), such as lane keeping and braking assistants, in order to obtain a fully autonomous driverless car. The Society of Automotive Engineers (SAE) provides, in fact, six possible levels of autonomy, from \emph{level 0}, with no assistance, to \emph{level 5}, where the presence of the driver inside the car is not needed at all. 

All these innovations have a common denominator: \emph{information technology}. Current top end vehicles have about 200 million lines of code, up to 200 Electronic Control Units (ECUs) and more than 5 km copper wires \cite{Simacsek2019}, which means cars are becoming very complex software-based IT systems. This fact marks a significant shift in the industry: the \emph{"mechanical"} world of original equipment manufacturers (OEMs) is converging towards that of IT companies.
\newline
In this context, the safety of modern vehicles is strictly related to addressing cybersecurity challenges. The electronic architecture of the vehicle has been designed and standardised over the years as a \emph{"closed"} system, in which all the data of the ECUs persisted in the internal network. The above new services require instead that data spread across multiple networks; 
there is, therefore, a bigger \emph{attack surface}, \ie new possibilities to be vulnerable to the attackers. Hence, automotive OEMs need to reinvent the car architecture with a \emph{secure-by-design} approach.

Another implication of this transformation is that the vehicle will be a fully-fledged cyber-physical system (CPS), that is \textquote{a system of collaborating computational elements controlling physical entities} \cite{Minerva2015}.
This definition reminds that, in terms of security, both the \emph{cyber}- and the \emph{physical}-related aspects should be considered. As an example, an autonomous car heavily interacts with the real world environment and faces the challenge of guaranteeing the resilience of the sensing and actuation devices. Therefore, security in automotive also involves addressing the specific issues of a CPS, as can be read in the work by \textcite{Wang2010}; however, in this paper, we will consider the attacks that are carried out in the cyber-space. In particular, we propose a systematisation of knowledge that focuses on the in-vehicle network, with the aim to provide the core elements for further analyses about complementary aspects of automotive cybersecurity.

\textbf{Paper structure.} In this paper, Section~\ref{sec:2} firstly lists the constraints in car design, then describes the principal standards for the internal network and the related security vulnerabilities. Section~\ref{sec:3} presents the various goals of the cyberattacks against vehicles, while Section~\ref{sec:4} makes an overview of the attack strategies. Section~\ref{sec:5} illustrates the proposed solutions for new architectures. Finally, Section~\ref{sec:6} discusses how the security evaluation can be expandend to address the impact of artificial intelligence and V2X, and Section~\ref{sec:conclusion} makes concluding remarks.
 \section{Automotive Networks}
\label{sec:2}

This Section describes the basic characteristics of a car internal network, from the design constraints to the main protocols and their vulnerabilities.

	\subsection{Constraints}
	\label{sec:21}

Although common IT security concepts can be used to design car electronics, there are some specific constraints to consider both in the hardware and the software side, as summarised by \textcite{Studnia2013} and \textcite{Pike2017}:
\begin{description}
	\item[Hardware limitations] The typical ECUs for cars are embedded systems with substantial hardware limitations, that is with low computing power and memory. This restriction means some security solutions like cryptography might be not fully implementable.
    Moreover, the ECUs are exposed to demanding conditions (such as low/high temperatures, shocks, vibrations, electromagnetic interferences), and must have an impact on the size and weight of the vehicle as small as possible. This is why the bus topology, which requires a much lower number of wires, is preferable compared to the star one.
    \newline
    These constraints cause the OEMs to be sensitive to component costs, which limits the possibility to embrace innovations.
	\item[Timing] Several ECUs must perform tasks with fixed real-time constraints, which are often safety-critical. Therefore, any security measure must not impact these tasks.
    \item[Autonomy] Since the driver must be focused on driving, the car should be as much autonomous as possible when protection mechanisms take place.
    \item[Life-cycle] The life-cycle of a car is much longer than that of conventional consumer electronics, so the need for durable hardware and easy-to-update software (especially security-related one).
    \item[Supplier Integration] To defend intellectual property, suppliers often provide (software) components without source code; therefore, any modification to improve security can be more difficult.
\end{description}

	\subsection{Main Standards}
	\label{sec:22}

Current vehicles mix different types of networks to let the dozens of ECUs communicate. The primary standards, typically suited for a specific domain and the related requirements, are: LIN, MOST, FlexRay and CAN; the latter represents the backbone of the entire network, so it is the most explanatory protocol to understand the critical points in automotive cybersecurity. It is worth noting that, due to the transitioning phase in the industry, the topology and the standards are going to change, as will be better illustrated in Section~\ref{sec:5}. Following the survey by \textcite{Huo2015}, the main features of CAN and Automotive Ethernet ---one of the new proposed protocols--- are the following:

\begin{figure}[htp]
\centering
\includegraphics[width=0.7\linewidth]{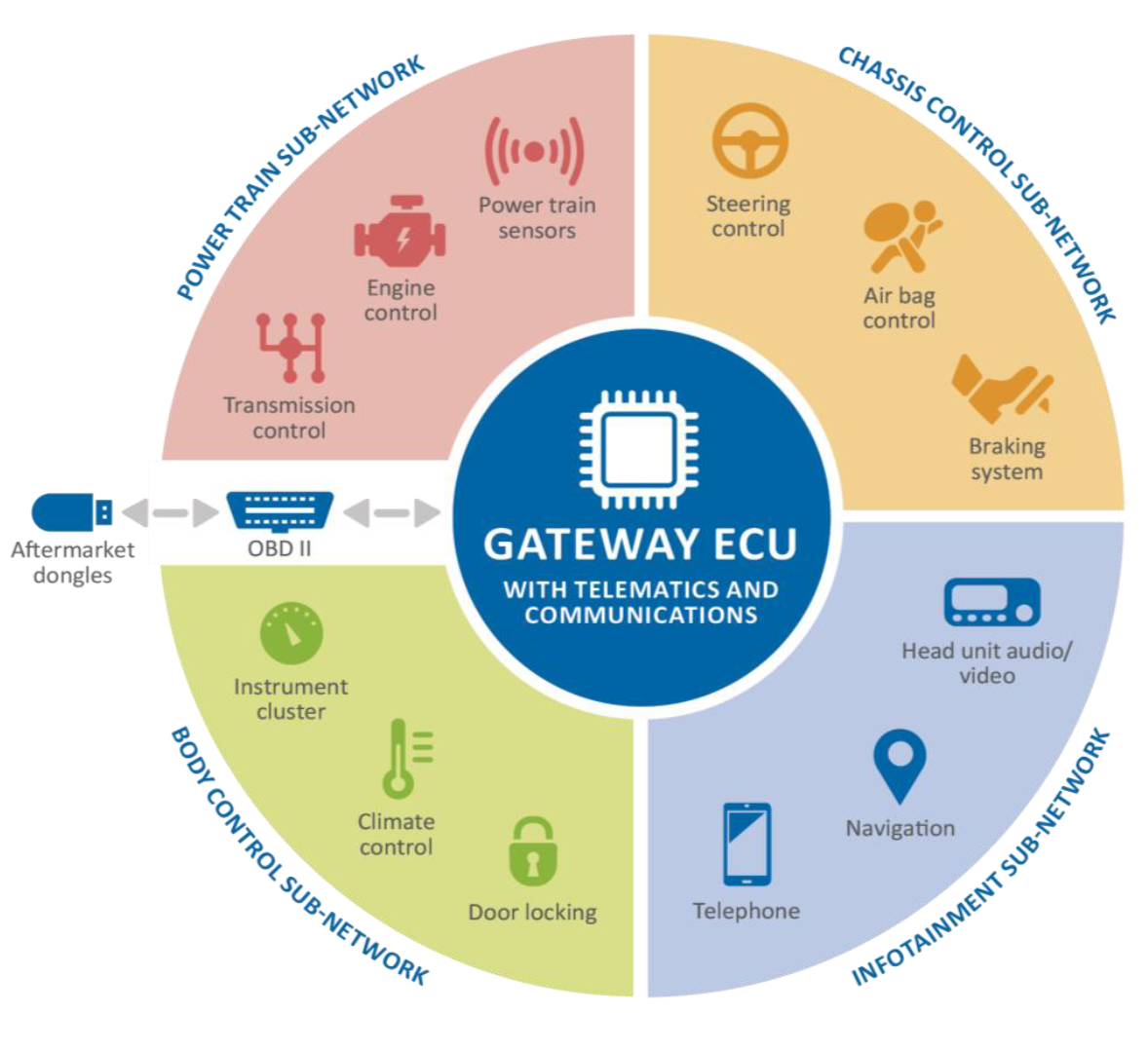}
\caption{Main domains in a modern car.~\autocite{ENISA2017}}
\label{fig:22domains}
\end{figure}

\begin{description}
\item[CAN] The Controller Area Network is the most used protocol for the in-vehicle network. It was released in 1986, but several variants and standards have been developed over the years. For simplicity, there is a low speed CAN that reaches up to 125 Kb/s while the high-speed version reaches up to 1 Mb/s; the first one it's suited for the body domain, the other one is used in \emph{'powertrain'} (engine or transmission control) and \emph{'chassis'} (suspension, steering or braking) domain. The CAN network is implemented with twisted pair wires, and an essential aspect is the network topology, which is a bus line. Although current designs are transitioning to a slightly different setting (Figure~\ref{fig:22domains}), with a Domain Controller (DC) that manages different sub-networks for each domain (\ie functionality), the main idea is still that the CAN bus acts as the backbone and all the data spread across the entire network, in broadcast mode.
\item[Automotive Ethernet] Although its adoption is still limited, Ethernet has a crucial role for next-generation automotive networks; it is a widespread standard for the common IT uses, and its high bandwidth is a desirable characteristic for modern vehicles. However, as it is, its cost and weight are not suited for automotive, hence the need for \emph{'Automotive Ethernet'}: in the past few years, among the various proposals, the \emph{'BroadR-Reach'} variant by Broadcom emerged and now its scheme has been standardised by IEEE (802.3bp and 802.3bw); moreover, other variants are under development by ISO. The standard is currently guided by the One-Pair Ether-Net (OPEN) alliance.
\newline
The main difference compared to standard Ethernet is the use of a unique unshielded twisted pair, which let the cost, size and weight significantly decrease, without sacrificing the bandwidth (100 or 1000 Mb/s).

\end{description}

Before moving on to the description of the vulnerabilities caused by these designs, it is useful to introduce an essential standard for diagnostics: OBD. It stands for On-Board Diagnostics, and it consists in a physical port, mandatory for US and European vehicles, that enables self-diagnostic capabilities in order to detect and signal to the car owner or a technician the presence of failures in a specific component. It gives direct access to the CAN bus, then causing a serious security threat, as will be described in Section~\ref{sec:4}; moreover, anyone can buy cheap dongles for the OBD port, extract its data and read them for example with a smartphone app.
	
	\subsection{Vulnerabilities}
    \label{sec:23}

The constraints described in Section~\ref{sec:21}, such as the need to reduce the cost and the size impact of the network, together with the past context in which the in-vehicle data was not exposed to external networks, caused the presence in the (CAN) backbone of the following design vulnerabilities \autocite{Liu2017}:

\begin{description}
\item[Broadcast transmission] Because of the bus topology, the messages between the ECUs spread across the entire network, causing a severe threat: accessing one part of the network (for example the OBD port) implies the possibility to send messages to the entire network or being able to eavesdrop on these communications.
\item[No authentication] There is no authentication that indicates the source of the frames, which means it is possible to send fake messages from every part of the network.
\item[No encryption] The messages can be easily analysed or recorded in order to figure out their function.
\item[ID-based priority scheme] Each CAN frame contains an identifier and a priority field; the transmission of a high priority frame causes the lower priority ones to back off, which enables Denial of Service (DoS) attacks.
\end{description}
 \section{Attack Goals}
\label{sec:3}

In this Section, different motivations that attract the attackers are described. Taking the works by \textcite{Studnia2013} and \textcite{IET2014} as references, these are the possible attack goals:
\begin{description}
\item[Vehicle theft] This is a straightforward reason to attack a vehicle.
\item[Vehicle enhancement] This refers to software modifications especially realised by the owner of the car. The goal might be to lower the mileage of the vehicle, tune the engine settings or install unofficial software in the infotainment.
\item[Extortion] This can be achieved for example through a ransomware-like strategy, \ie blocking the victim's car until a fee is paid.
\item[Intellectual challenge] The attack is conducted to demonstrate hacking ability.
\item[Intellectual property theft] This refers to the elicitation of the source code for industrial espionage.
\item[Data theft] This is an increasingly important goal, a consequence of the new paradigm of connected cars. There are different types of data to steal, such as:
	\begin{itemize}
	\item License plates, insurance and tax data;
    \item Location traces;
    \item Data coming from the connection with a smartphone, such as contacts, text messages, social media data, banking records.
	\end{itemize}
The combination of these data might allow the attacker to discover the victim's habits and points of interest, exposing him to burglary or similar attacks.
\end{description}
 \section{Attack Scenarios}
\label{sec:4}

In this Section, an overview of attack techniques and examples is provided. Following the work by \textcite{Liu2017}, the typical attack scheme includes an initial phase in which a physical (\eg OBD) or wireless (\eg Bluetooth) car interface is exploited in order to access the in-vehicle network. The most common interface to access it is OBD, but several works leverage different entry points: \textcite{Checkoway2011} succeeded in sending arbitrary CAN frames through a modified WMA audio file burned onto a CD. \textcite{Mazloom2016} showed some vulnerabilities in the MirrorLink standard that allow controlling the internal CAN bus through a USB connected smartphone. \textcite{Rouf2010} analysed the potential vulnerabilities in the Tire Pressure Monitoring System (TPMS), while \textcite{Garcia2016} found out that two widespread schemes for keyless entry systems present vulnerabilities that allow cloning the remote control, thus gaining unauthorised access to the vehicle. 

Once the interface is chosen, then the following methodologies are used to prepare and implement the attack:
\begin{description}
\item[Frame sniffing] Leveraging the broadcast transmission and the lack of cryptography in the network, the attacker can eavesdrop on the frames and discover their function. It is the typical first step to prepare the attack. An example of CAN frames sniffing and analysis is the work by \textcite{Valasek2013}.
\item[Frame falsifying] Once the details of the CAN frames are known, it is possible to create fake messages with false data in order to mislead the ECUs or the driver, \eg with a wrong speedometer reading.
\item[Frame injection] The fake frames, set with a proper ID, are injected in the CAN bus to target a specific node; this is possible because of the lack of authentication.
An illustrative ---and very notorious--- attack regards the exploitation made by \textcite{Miller2015} towards the 2014 Jeep Cherokee infotainment system, which contains the ability to communicate over Sprint's cellular network in order to offer in-car Wifi, real-time traffic updates and other services. This remote attack allowed to control some cyber-physical mechanisms such as steering and braking. The discovery of the vulnerabilities in the infotainment caused a 1.4 million vehicle recall by FCA.

\item[Replay attack] In this case, the attacker sends a recorded series of valid frames into the bus at the appropriate time, so he can repeat the car opening, start the engine, turn the lights on. \textcite{Koscher2010} implemented a replay attack in a real car scenario.
\item[DoS attack] As anticipated in Section~\ref{sec:23}, flooding the network with the highest priority frames prevents the ECUs from regularly sending their messages, therefore causing a denial of service. An example of this attack is the work by \textcite{Palanca2017}.

\end{description}
 \section{Security Countermeasures}
\label{sec:5}

This Section firstly aims to summarise the basic security principles to consider when designing car electronics and related technology solutions. Then, it focuses on the major projects for new architectures.

	\subsection{Requirements}
    \label{sec:51}
    
A typical pattern to help to develop secure architectures is the so-called \emph{'CIA triad'}, \ie three conditions that should be guaranteed as far as possible; they are: \emph{confidentiality}, \emph{integrity}, \emph{availability}. As the previous Sections demonstrated, none of them is inherently guaranteed through the current reference backbone ---the CAN bus.
\newline
Bearing in mind these concepts and taking a cue from the work by \textcite{ACEA2017}, the proposed countermeasures and some of the related implementations in the research literature are the following:
\begin{description}
\item[Dedicated HW] To supply the scarcity of computing power of the ECUs and satisfy the real-time constraints, it may be necessary to integrate hardware platforms specifically designed for security functions. This approach has been pursued, for example, in the EVITA and HIS project, and it is referred to as Hardware Security Module (HSM) or Security Hardware Extension (SHE).
\item[Cryptography] Encryption can help in ensuring confidentiality and integrity. It is worth noting that implementing cryptography is not trivial, since the low computing power may prevent the OEMs from using robust algorithms, which means cryptography might be even counter-productive. The guidelines recommend state-of-the-art standards, taking care of key management and possibly using dedicated hardware. There are several works about cryptography; for example, \textcite{Zelle2017} investigated whether the well-known TLS protocol applies to in-vehicle networks.
\item[Authentication] Since different ECUs interact with each other, it is fundamental to know the sender of every incoming message. Two recent works that integrate authentication are those by \textcite{Mundhenk2017auth} and \textcite{VanBulck2017}.
\item[Access control] Every component must be authorised in order to gain access to other parts. The guidelines suggest adopting the principle of least privilege, \ie a policy whereby each user (each ECU in this case) should have the lowest level of privileges which still permits to perform its tasks.
\item[Isolation/Slicing] This hardening measure aims at preventing the chance for an attacker to damage the entire network. This goal can be achieved for example isolating the driving systems from the other networks (\eg the infotainment), or through a central gateway that employs access control mechanisms.
\item[Intrusion detection] Intrusion Detection Systems (IDSs) monitor the activities in the network searching for malicious or anomalous actions. Some examples in the literature are the works by \textcite{Song2016} and by \textcite{Kang2016}, which uses deep neural networks.
\item[Secure updates] The Over-The-Air (OTA) updates are on the one hand a risk that increases the attack surface; on the other, they are an opportunity to quickly fix the discovered vulnerabilities (besides adding new services). Some recent works to secure the updates but also V2X communications are those by \textcite{Dorri2017} and \textcite{Steger2018}, both taking advantage of blockchain.
\item[Incident response and recovery] It is necessary to ensure an appropriate response to incidents, limit the impact of the failures and be always able to restore the standard vehicle functionality.
\end{description}

All the above aspects should be fulfilled in a Security Development Lifecycle (SDL) perspective, with data protection and privacy as a priority. Testing and information sharing among industry actors are recommended.

	\subsection{Main Projects}
	\label{sec:52}

\begin{figure}[htp]
\centering
\includegraphics[width=\linewidth]{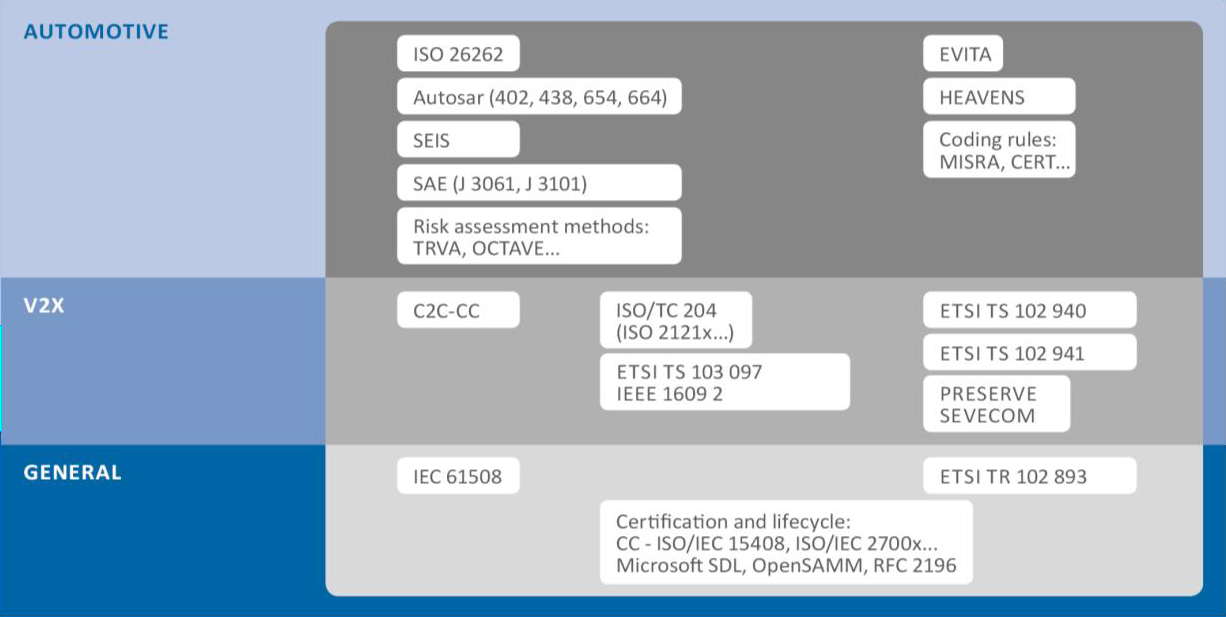}
\caption{Safety and security initiatives inside and outside of the automotive domains. (\textcite{ENISA2017})} 
\label{fig:52initiatives}
\end{figure}

In the past ten years, several research proposals and standardisation projects started, aiming to develop and integrate the ideas of the previous Section organically; a map of these initiatives can be seen in Figure~\ref{fig:52initiatives}.

Among these projects, SAE J3061\footnote{\url{https://www.sae.org/standards/content/j3061_201601/}}, finalised in 2016, guides vehicle cybersecurity development process, ranging from the basic principles to the design tools. However, a new international standard, the ISO/SAE 21434, is under development; its goal is to \begin{enumerate*}[label={(\alph*)}, font=\itshape]
\item describe the requirements for risk management
\item define a framework that manages these requirements, without indicating specific technologies, rather giving a reference, useful also for legal aspects.
\end{enumerate*}

Moreover, the implementation of these guidelines and the transition towards a new in-vehicle network architecture is currently guided by some projects like AUTOSAR\footnote{\url{https://www.autosar.org}}. This initiative is a partnership born in 2003 between several stakeholders, ranging from the OEMs to the semi-conductors companies, which aims to improve the management of the E/E architectures through reuse and exchangeability of software modules; concretely, it standardises the software architecture of the ECUs. It is still an active project, now also focused on autonomous driving and V2X applications, and it covers different functionalities, from cybersecurity to diagnostic, safety, communication. AUTOSAR also supports different software standards, such as GENIVI\footnote{\url{https://www.genivi.org}}, another important alliance aiming to develop open software solutions for In-Vehicle Infotainment (IVI) systems.
 \section{Discussion}
\label{sec:6}

\begin{figure}[htp]
\centering
\includegraphics[width=\linewidth]{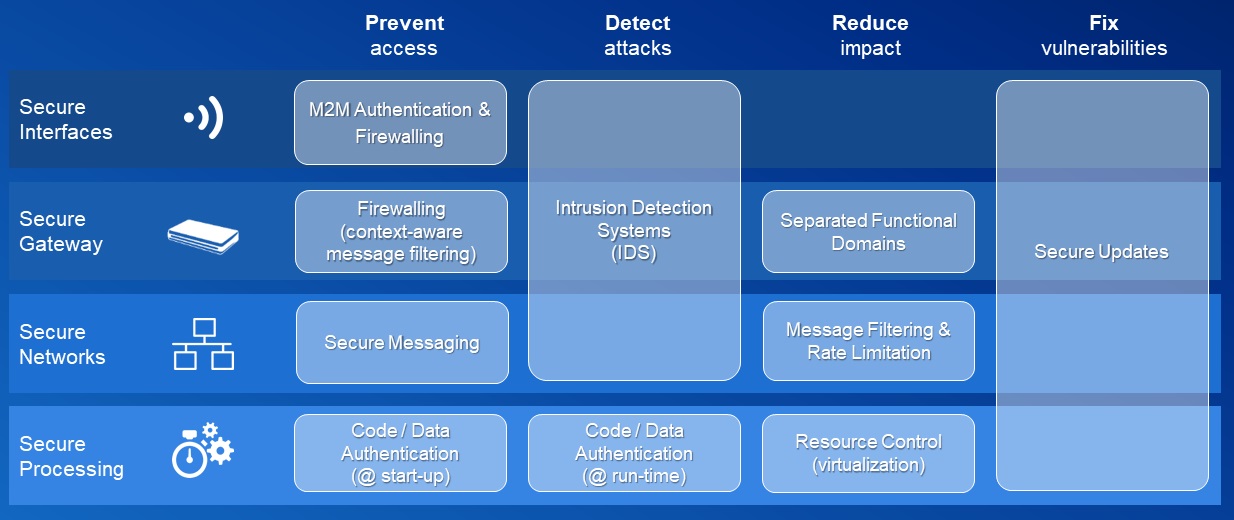}
\caption{Applying security principles (\cite{Simacsek2019})} 
\label{fig:53secprinciples}
\end{figure}

The ideas expressed in the previous Section can be summarised by Figure~\ref{fig:53secprinciples}, which shows how the security principles can be implemented in practice. In our opinion, the primary protocol upon which the backbone of the future in-vehicle network will be built is Automotive Ethernet. Moreover, the takeaway message from these initiatives is the specific focus on security: each building block implies a research activity aimed at proposing a solution tailored for the automotive domain.

In this paper, we examined the core elements and concerns for secure internal networks; however, it is worth discussing, although in an introductory manner, about how the same awareness should be extended to the very new actors in automotive, \ie artificial intelligence and V2X. These elements enable new advanced, \emph{smart} services ---\eg \emph{platooning}, that is the use of a fleet of vehicles that travel together in a coordinated and autonomous way--- and, as a consequence, further threats. In particular, focusing on artificial intelligence, the primary concerns come from autonomous driving, where deep learning is the main enabling technology. In addition to the inherent complexity in developing a fully autonomous car for the real world, several studies demonstrated how machine learning-based algorithms are vulnerable, i.e. the fact that carefully-perturbed inputs can easily fool classifiers, causing, for example, a stop sign to be classified as a speed limit (\cite{Gu2019}). These issues originate the research topic of \emph{adversarial learning}.  Moreover, the use of machine learning is not limited to computer vision but also includes cybersecurity software, such as IDSs, and safety systems, such as drowsiness and distraction detectors. Therefore, it is fundamental to leverage proper techniques (\eg \cite{Demontis2017}) to 
\begin{enumerate*}[label=\textit{\alph*)}]
    \item avoid consistent drops of performances
    \item increase the effort of the attacker to evade the classifiers
    \item keep the complexity of the algorithms within an acceptable level, given the constraints described in Section~\ref{sec:21}.
\end{enumerate*}
Ultimately, these concerns must be addressed with the same attention as the ones related to the internal network architecture. In this sense, some works, such as \cite{Salay2017}, propose to include machine learning-specific recommendations in the ISO 26262~\footnote{\url{https://www.iso.org/standard/68383.html}} standard.
 \section{Conclusion}
\label{sec:conclusion}

To sum up, in this paper we deduced how the digitalisation process within the automotive industry, where the OEMs are converging towards IT companies and the vehicles are becoming "smartphones on wheels", came up against serious cybersecurity issues, due to security flaws inherited by an original design where the in-vehicle network did not interact with the external world. By contrast, the Mobility-as-a-Service paradigm causes the vehicle to be hyper-connected and consequently much more exposed to cyber threats.

In this transition phase, we observed the effort in developing more and more complex platforms in a safety-critical context with strict requirements such as the limited hardware and the real-time constraints. For these reasons, both the industry and the researchers are pledging to leverage the common IT methodologies from other domains and tailor them for the automotive one. The route towards this goal is not straightforward, as noted in the study by \textcite{Ponemon2019}: 84\% of the professionals working for OEMs and their suppliers still have concerns that cybersecurity practices are not keeping pace with evolving technologies.

As a final remark, we claim that the core ideas concerning the in-vehicle network, and described in this paper, could be considered for further analyses on the security for autonomous driving and V2X communications.

\section*{Acknowledgement}

The authors thank Abinsula srl for the useful discussions on the mechanisms of the automotive industry and its trends.
 \printbibliography

\end{document}